**Title**: Longitudinal abnormalities in white matter extracellular free water volume fraction and neuropsychological functioning in patients with traumatic brain injury

**Authors**: James J Gugger,*[1] Alexa E Walter,[1] Drew Parker,[2,3,4] Nishant Sinha,[1] Justin Morrison,[1] Jeffrey Ware,[2] Andrea LC Schneider,[1,5] Dmitriy Petrov,[3] Danielle K Sandsmark,[1] Ragini Verma,**[2,3,4] Ramon Diaz-Arrastia**[1]

* Corresponding author: James J Gugger

** Both authors contributed equally to this work.

Affiliations:

[1]Department of Neurology, University of Pennsylvania Perelman School of Medicine, Philadelphia, Pennsylvania

[2]Department of Radiology, University of Pennsylvania Perelman School of Medicine, Philadelphia, Pennsylvania

[3]Department of Neurosurgery, University of Pennsylvania Perelman School of Medicine, Philadelphia, Pennsylvania

[4]Diffusion and Connectomics in Precision Healthcare Research Lab, University of Pennsylvania Perelman School of Medicine, Philadelphia, Pennsylvania

[5]Department of Biostatistics, Epidemiology, and Informatics, University of Pennsylvania Perelman School of Medicine, Philadelphia, Pennsylvania

**Abstract**

Traumatic brain injury is a global public health problem associated with chronic neurological complications and long-term disability. Biomarkers that map onto the underlying brain pathology driving these complications are urgently needed to identify individuals at risk for poor recovery and to inform design of clinical trials of neuroprotective therapies. Neuroinflammation and neurodegeneration are two endophenotypes associated with increases in brain extracellular water content after trauma. The objective of this study was to describe the relationship between a neuroimaging biomarker of extracellular free water content and the clinical features of patients with traumatic brain injury. We analyzed a cohort of 64 adult patients requiring hospitalization for non-penetrating traumatic brain injury of all severities as well as 32 healthy controls. Patients underwent brain MRI and clinical neuropsychological assessment in the subacute (2-weeks) and chronic (6-months) post-injury period, and controls underwent a single MRI. For each subject, we derived a summary score representing deviations in whole brain white matter (1) extracellular free water volume fraction (VF) and (2) free water-corrected fractional anisotropy (fw-FA). The summary specific anomaly score (SAS) for VF was significantly higher in TBI patients in the subacute and chronic post-injury period relative to controls. SAS for VF significantly correlated with neuropsychological functioning in the subacute, but not chronic post-injury period. These findings indicate abnormalities in whole brain white matter extracellular water fraction in patients with TBI and are an important step toward identifying and validating noninvasive biomarkers that map onto the pathology driving disability after TBI.

**Introduction**

Traumatic brain injury (TBI) is a global public health issue where even its most mild form—mild TBI or concussion—there is a substantial risk for chronic post-traumatic symptoms and long-term disability.[1] Neurotrauma results in injury to multiple brain structures including neurons, glia, and cerebral blood vessels with considerable person-to-person variability in terms of severity and anatomic location of injury. This mechanistic and spatial heterogeneity likely underlies the interindividual variability in outcomes post-injury. Abnormalities seen on clinical neuroimaging such as subarachnoid hemorrhage and contusions may improve long-term outcome prediction after TBI,[2,3] but do not explain enough of the variance to be useful tools for risk stratification. Quantitative tools, such as advanced neuroimaging, offer promise for providing more accurate risk stratification after TBI.

One such advanced neuroimaging tool used extensively in TBI studies is diffusion tensor imaging (DTI) obtained from diffusion-weighted MRI. DTI models white matter microstructural integrity with variables such as fractional anisotropy (FA) and mean diffusivity (MD).[4,5] Multiple studies have demonstrated diffuse white matter abnormalities, namely decreases in FA and increases in MD, across the TBI severity spectrum.[6] Although primarily thought to reflect diffuse axonal injury, DTI models water diffusion, which could be abnormal for a multitude of reasons, including axonal injury, gliosis, atrophy, edema, and partial volume effects at brain/CSF boundaries.[7] Given that all of these processes occur in TBI, novel methods for characterizing white matter microstructure are needed to improve the specificity of quantitative neuroimaging biomarkers with the goal of creating a clinical decision making tool in TBI. One such method is free water correction,[8] which models water diffusion at the extracellular (isotropic) level. Water diffusion is more isotropic in the presence of CSF and edema; and free water correction can map this at the voxel level while also providing voxel-wise estimates of traditional DTI variables (e.g., FA) corrected for extracellular free water. These corrected DTI variables may provide a more accurate assessment of microstructural abnormalities within CNS cells (e.g., axonal injury).

A major challenge with DTI and other quantitative neuroimaging modalities is that there are few robust measures for summarizing abnormalities occurring with variable spatial distribution across brain. This is of particular concern for brain disorders characterized by pathologically heterogenous abnormalities such as TBI. To facilitate data reduction, voxel-wise estimates are typically averaged across neuroanatomically meaningful regions of the brain (regions of interest [ROI]) that correspond to named anatomical structures (e.g., white matter tracts, cortical gyri and sulci, subcortical nuclei). This approach generates hundreds of variables that provide little useful information when analyzed with univariate statistics where values are averaged across a cohort of patients with heterogenous brain pathology. One potential solution to this problem is to generate a single composite measure summarizing imaging abnormalities across the entire brain using the Mahalanobis distance, a multivariate generalization of the z-score.[9] This approach was originally proposed as a method for describing deviation from normal structural connectivity in mild TBI. [9]

In a recent study by our group, we validated a Mahalanobis distance based summary score of free water VF as a measure of injury severity in moderate-to-severe TBI.[10] This measure was associated with worse long-term executive functioning and processing speed. Similarly, in a study of cognitively diverse individuals (without TBI), free water volume fraction correlated with episodic memory and executive function.[11] The main objective of this study was to assess the utility of a summary score based on the Mahalanobis distance to describe anomalies in white matter microstructure and explore its correlation with neuropsychological functioning in an independent cohort of patients with primarily mild TBI.

**Materials and Methods**

*Participants*

In this single center observational study, we recruited adult patients with non-penetrating TBI who required hospitalization post-injury. Patients were enrolled at the time of hospital admission. Patients were excluded for a history of pre-existing serious neurological or psychiatric disorder, comorbid disabling condition limiting outcome assessment, current pregnancy, if they were incarcerated, or had low MRI image quality. Patients with focal intraparenchymal lesions exceeding 50 cm$^3$ were also excluded.[10,12] Patients underwent brain MRI in the subacute and chronic post-injury periods. We also enrolled healthy control subjects recruited from the general population with no history of TBI within one year of enrollment, pre-existing disabling neurological or psychiatric disorder, or current pregnancy. Healthy controls underwent a single brain MRI. Demographic information, medical history, admission injury characteristics, and other clinical information were collected from the medical record. This study was approved by the Institutional Review Board at the University of Pennsylvania.

*Neuropsychological assessment*

Patients underwent neuropsychological assessment by trained research personnel, who were blind to imaging findings. Assessments were performed on the same day as the MRI at approximately two weeks and six months post-injury. Neuropsychological function was assessed across three domains: processing speed assessed with the Processing Speed Index from the Wechsler Adult Intelligence Scale IV;[13] verbal learning by Rey Auditory Verbal Learning Test (RAVLT);[14] and executive functioning by the Trail Making Test-Parts A and B.[15]

*Image acquisition*

Brain MRIs were performed on a 3T scanner (Siemens Prisma) using a product 32-channel head coil. Structural imaging included a sagittal T1-weighted MPRAGE (TR = 2.3 s, TE = 2.94 ms, TI = 900 ms, FA = 9°, resolution = 1 x 1 x 1 mm). Whole brain diffusion MRI (dMRI) was performed with an echo planar sequence with FA 90° and resolution = 2.4 x 2.4 x 2.4 mm (b-value = 1000 s/mm$^2$, 64 diffusion directions, TR 2.9 s, TE 94 ms). Separately, 14 images with b-value = 0 s/mm$^2$ with reverse phase encoding were acquired.

*Image processing*

Structural MRI and dMRI data were visually inspected for artifacts before preprocessing. T1-weighted images were bias corrected using the N4BiasCorrection tool from ANTs[16], followed by brain extraction using a multi-atlas segmentation tool, MUSE[17]. Diffusion MRI data was preprocessed in three steps: First, local-PCA denoising[18] as implemented in MATLAB 2021a (MathWorks, Natick, MA); followed by motion and distortion correction using the top up and eddy tools in FSL;[19] then bias field correction using N4BiasCorrection from ANTs. A tensor model[4] was fit to the data in DIPY.[20] We used FERNET (Freewater EstimatoR using iNtErpolated iniTialization),[8] a recent method for free water correction in clinically acquired DTI data, to estimate the free water-volume fraction (VF) and free water-corrected FA (fw-FA). An example case of a VF map is shown in Figure 1. Using the deformable SyN registration algorithm from ANTs,[21] the data were registered to the JHU-MNI-ss (Eve) template.[22] Mean values of VF and fw-FA were calculated in 92 regions of interest (ROIs) of the Eve white matter parcellation.

Fornix and tapetum regions were excluded because of their small size and proximity to the ventricles. For each ROI and each DTI variable (i.e., VF, fw-FA), a linear regression was run using only observations from controls with age as the independent variable and the mean of the respective DTI variable (e.g., VF) as the dependent variable. Age was then regressed out for both patients and controls by obtaining the residuals ($X^r_i$) for each ROI from the above regression.

*Patient-specific summary anomaly score of white matter microstructure*

For each DTI measure, we defined a summary specific anomaly score (SAS) for each participant using the Mahalanobis distance, a multivariate generalization of the z-score:

$$M = \sqrt{(s-\mu)^T \times C^{-1} \times (s-\mu)}$$

Where s = [$X^r_1$, $X^r_2$... $X^r_{92}$] is a 92x1 vector containing the residuals of the age-regressed DTI variable ($X^r_i$) for each ROI for the participant; μ is a 92x1 vector of mean $X^r$ of each ROI in the healthy controls; and C is the covariance matrix between ROIs across the control population. Thus, for each patient and control we calculated M, the Mahalanobis distance-based SAS for each dMRI measure: VF and fw-FA. This is a validated method for describing whole brain abnormalities in TBI and is described in more detail elsewhere.[9,10]

Calculation of Mahalanobis distance requires that the number of reference observations (i.e., control subjects) exceed the number of variables (i.e., ROIs). Since the number of white matter ROIs in the Eve atlas was larger than the number of controls in our cohort, we combined the controls from our cohort with 77 control subjects enrolled in the Transforming Research and Clinical Knowledge in Traumatic Brain Injury (TRACK-TBI) study (mean age (SD) of 39 (15.0); 51 men (66%)). dMRI data from controls from TRACK-TBI and controls from our cohort were harmonized using ComBat,[23] a batch effect removal technique that removes acquisition and processing differences while retaining the effects of biology (e.g., age, sex). Thus, the SAS represents an age- and sex-corrected measure of whole brain white matter microstructural anomaly referenced to a cohort of 109 healthy controls. We derived SAS separately for VF and fw-FA for each control (from our cohort) and at each time point for each patient. We limited all subsequent analyses to patients and controls enrolled at our center and did not calculate SAS for controls from TRACK-TBI.

*Statistical analysis*

The nonparametric Mann Whitney U test was used for comparison of independent continuous data whereas the Wilcoxon signed rank test was used for dependent continuous data. Chi-squared or Fisher's exact test were used for categorical data where appropriate. Cohen's *d* was computed to estimate effect size of the SAS in patients relative to controls. We used principal component analysis (PCA) to derive a measure of overall cognitive functioning suitable for comparison with the summary measure of whole brain white matter microstructural integrity. We performed two separate PCAs corresponding to neuropsychological assessments completed in the subacute and chronic post-injury period. Each PCA included total z-scores for each of the four neuropsychological assessments and the first principal component score was extracted for further analysis. Using the first principal component score at each time point as a proxy for overall neurocognitive status, we assessed the relationship between neuropsychological functioning and SAS using Spearman's rank correlation. Results were statistically significant if the p-value was less than 0.05. We repeated the above analysis including only patients with mild TBI (i.e., GCS 13-15) to ensure that findings were not driven by patients with more severe injury (excluded 5 participants with moderate or severe injury). All analyses were carried out using MATLAB 2022a (MathWorks, Natick, MA).

Results

*Demographic and clinical information*

We enrolled 70 patients and 32 controls. Two patients were excluded due to low MRI quality or artifact. Four patients with focal intraparenchymal lesions exceeding 50 cm$^3$ were excluded. Demographic and clinical information for the cohort is presented in Table 1. Injury severity according to GCS category was predominantly mild (89%); Five patients had moderate or severe TBI (8%). An acute intracranial lesion was visible on CT in 70% of patients.

*Summary specific anomaly score in the subacute and chronic post-injury period*

Fifty-nine patients underwent brain MRI in the subacute post-injury period (median 17 days post-injury; interquartile range (IQR) 8-27 days) and 35 were scanned in the chronic post-injury period (median 195 days post-injury; IQR 169-220 days). Thirty patients underwent brain imaging at both time points. Figure 2 shows the summary specific anomaly score (SAS) for controls and patients in the subacute and chronic post-injury period for VF and fw-FA, respectively. SAS was significantly higher in TBI patients in the subacute post-injury period relative to controls for VF ($p < 0.00001$, Cohen's d = 0.9), but not fw-FA ($p = 0.08$, Cohen's d = 0.4). Similarly, SAS was significantly higher in TBI patients in the chronic post-injury period relative to controls for VF ($p = 0.0002$, Cohen's d = 1.0), but not fw-FA ($p = 0.14$, Cohen's d = 0.3). Among TBI patients, SAS remained stable at a group level from the subacute to chronic post-injury period for both VF and fw-FA. SAS for VF increased over time in 16 participants with TBI and decreased in 14; SAS for fwFA increased over time in 12 and decreased in 18. Sensitivity analyses performed among patients with GCS from 13-15 showed similar results (excluding 5 patients with moderate or severe TBI).

***Relationship between summary specific anomaly score and neuropsychological functioning***

Forty-nine patients had complete neuropsychological testing results available for analysis in the subacute post-injury period and 25 in the chronic post-injury period. To derive a summary measure of neuropsychological functioning suitable for comparison with SAS, we performed principal component analysis on the z-scores for Processing Speed Index from the Wechsler Adult Intelligence Scale IV, Rey Auditory Verbal Learning Test, and Trail Making Test-Parts A and B. As a proxy for overall neurocognitive functioning, we extracted the first principal component (PC1), which explained 68% and 57% of the variance for the two week and six-month measures, respectively. Results of the correlation between PC1 and the SAS for VF and fw-FA are presented in Figure 3. SAS for VF was moderately correlated with PC1 in the subacute post-injury period (Spearman rho = 0.41 and $p < 0.01$). Subacute fw-FA did not correlate with PC1 at 2 weeks post-injury. SAS for VF and fw-FA did not correlate with PC1 in the chronic post-injury period. Sensitivity analyses including only patients with GCS from 13-15 showed similar results.

**Discussion**

The main findings of this study are that a multivariate summary score describing abnormalities in VF distinguish patients with TBI from healthy controls in the subacute and chronic post-injury period and that this measure correlates with neuropsychological functioning in the subacute, but not chronic time-period. This study also provides proof of concept that a large database of controls can be harmonized to generate normative data for a summary statistic representing a patient-specific summary anomaly score.

TBI is heterogenous with respect to injury location as well as pathophysiology with contributions from axonal injury, vascular injury, and inflammation. Together, free water VF and fw-FA may be used for multivariate biophysical modeling of these dynamic processes. Free water volume fraction abnormalities may represent edema/inflammation or brain atrophy in this context whereas fw-FA may model the alterations in axonal coherence reflective of axonal injury. Understanding the burden of these abnormalities across the entire brain is important given the anatomic heterogeneity of TBI. Our results build upon prior studies of free water correction in patients with mild TBI. Palacios et al.[24] found increases in free water in a cohort of mild TBI using another biophysical model of water diffusion known as neurite orientation dispersion and density imaging. However, in this study free water decreased from the subacute to chronic post-injury period. Conversely, Pasternak et al.[25] found decreased free water from pre- to post-season in a cohort of subjects with sports-related concussion. Such differences may be explained by use of different biophysical models, different levels of injury severity, and different form of analysis (both studies used voxel-wise analyses). The results of this study are consistent with a prior study from our group on a cohort of patients with moderate-to-severe TBI. In our prior study, we found a stronger relationship between SAS for VF and neuropsychological functioning compared to fw-FA and FA from a standard single tensor model. Outside of TBI, free water imaging has been used to probe abnormal increase in free water in schizophrenia,[26] Parkinson's disease,[27,28] and dementia.[11,29]

We found large effect sizes for VF and smaller effect sizes for fw-FA that did not reach statistical significance. Given that VF models extracellular water diffusion and fw-FA models diffusion along the main axis after free water correction,[8] this suggests that the burden of axonal injury

(as modeled by fw-FA) in this cohort is less severe (and statistically similar to controls) than the burden of extracellular VF abnormalities. These findings are in agreement with our previous study of free water VF in moderate-to-severe TBI.[10] It should be noted that the underlying biological perturbations driving these changes cannot be resolved with neuroimaging. Based on pathologic data in TBI indicating chronic white matter degeneration and inflammation after TBI[30] we hypothesize that increases in free water in TBI may represent either inflammation/edema or atrophy. *Ex-vivo* correlation with histopathology is needed to confirm this hypothesis.

The summary specific abnormality score for VF correlated with neurocognitive function in the subacute post-injury period, but not in the chronic post-injury period. Brain pathology and cognition/function likely follow a dynamic trajectory that varies considerably from patient to patient.[31–33] Our interpretation of the findings from this cohort are that the physical manifestations of white matter injury are relatively stable from the subacute to chronic post-injury period, but that cognitive function tends to improve over this period. Although there were distinct groups of participants with either increases or decreases in SAS, the small sample size precluded meaningful analysis of the trajectory of SAS and how it maps to neurocognitive status.

This study has several limitations. First, the size of the cohort is relatively small, particularly at the chronic post-injury time point. Given the heterogeneity of brain pathology and recovery trajectory in TBI these findings require validation in larger cohorts such as TRACK-TBI. Second, we did not consider the effect of lesions on the SAS. Consistent with prior studies,[10,12] we excluded patients with large focal intraparenchymal lesions, but cannot resolve the effects that smaller lesions may have on SAS. We plan to address this issue in a future study using a lesion segmentation algorithm. Third, as mentioned above, free water is a biophysical model of voxel-wise water diffusion and is not specific to the underlying brain pathology.

*Conclusion*

We demonstrated the feasibility of generating a multivariate summary score using normative data to describe abnormalities in white matter extracellular free water volume fraction in patients with predominately mild TBI. We show that abnormalities in free water volume fraction are more severe than those thought to reflect axonal injury (i.e., fw-FA) and correlate with neuropsychological functioning in the subacute post-injury period. These findings are an important step towards identifying quantitative neuroimaging biomarkers that can be used in precision medicine clinical trials for risk-stratification in TBI.

**Acknowledgements**:  Work in the authors' laboratory was supported by the Pennsylvania Department of Health. This work was also funded by NINDS T32NS091006, U01NS086090, U01NS114140, DoD W81XWH-12-2-0139, W81XWH-14-2-0176, W81XWH-19-2-0002 and the American Epilepsy Society/Citizens United for Research in Epilepsy (Research and Training Fellowship for Clinicians).

**Table 1**: Demographic and clinical information

|  |  | Controls (n=32) | TBI patients (n=64) | p-value |
|---|---|---|---|---|
| Age (yrs., SD) |  | 30.7 (7.7) | 35.9 (16.2) | 0.089 |
| Male |  | 19 (59%) | 47 (73%) | 0.17 |
| High school education or greater |  | 26 (81%) | 54 (84%) | 0.77 |
| Injury severity by GCS category | Mild |  | 57 (89%) |  |
|  | Moderate |  | 3 (5%) |  |
|  | Severe |  | 2 (3%) |  |
| Injury cause | Road traffic incident |  | 32 (50%) |  |
|  | Fall |  | 18 (28%) |  |
|  | Other |  | 11 (22%) |  |
| Loss of consciousness |  |  | 44 (69%) |  |
| Post-traumatic amnesia |  |  | 37 (58%) |  |
| Prior TBI |  |  | 4 (6%) |  |
| Acute intracranial lesion on CT |  |  | 41 (64%) |  |
| Type of CT lesion |  |  |  |  |
|   Epidural hematoma |  |  | 3 (5%) |  |
|   Subdural hematoma |  |  | 19 (30%) |  |
|   Subarachnoid hemorrhage |  |  | 30 (47%) |  |
|   Non-specific extra-axial hemorrhage |  |  | 4 (6%) |  |
|   Intraventricular hemorrhage |  |  | 1 (2%) |  |
|   Contusion |  |  | 17 (27%) |  |
|   Diffuse axonal injury |  |  | 1 (2%) |  |

*Figure 1:* Representative case showing the white matter free water volume fraction map overlaid on the T1-weighted image of a participant with TBI scanned two weeks after injury. The color bar shows the free water volume fraction ranging from 0-1.

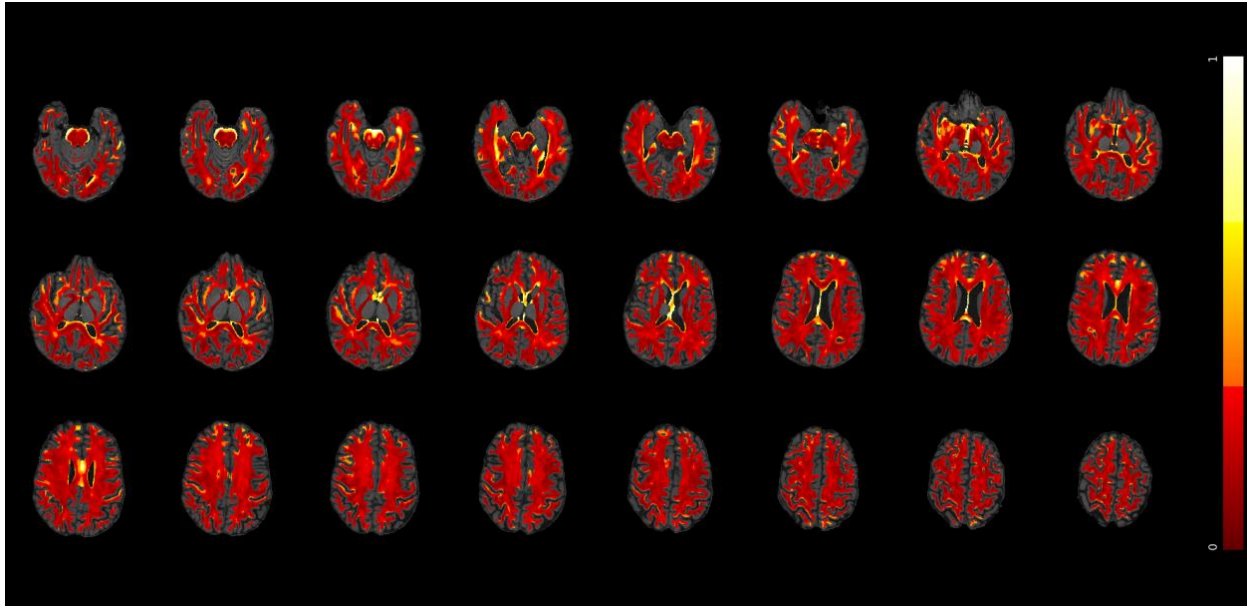

*Figure 2:* Summary specific anomaly score (SAS) for volume fraction (VF) and free water-corrected fractional anisotropy (fw-FA) for TBI patients and controls. Higher SAS indicates more white matter abnormalities.

* $p < 0.00001$, Cohen's d = 0.9 for comparison between SAS VF in the subacute post-injury period relative to controls

**$p = 0.0002$, Cohen's d = 1.0 for comparison between SAS VF in the chronic post-injury period relative to controls

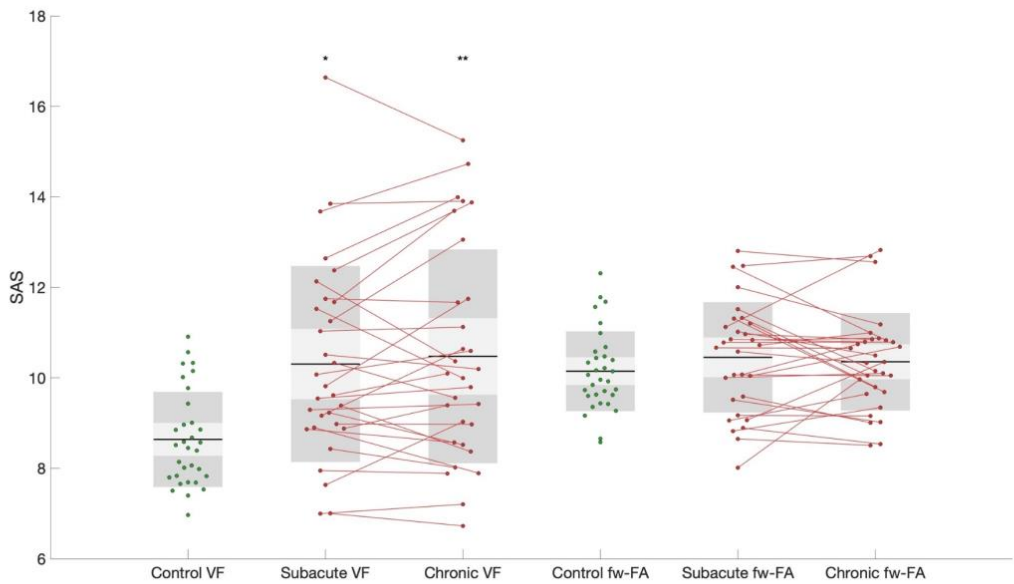

*Figure 3:* Association between summary anomaly scores (SAS) for DTI and neuropsychological function summarized by the first principal component (PC1) of the measures of processing speed, verbal learning, and executive function. Higher SAS indicates more white matter abnormalities. Higher PC1 score indicates worse performance on neuropsychological functioning.

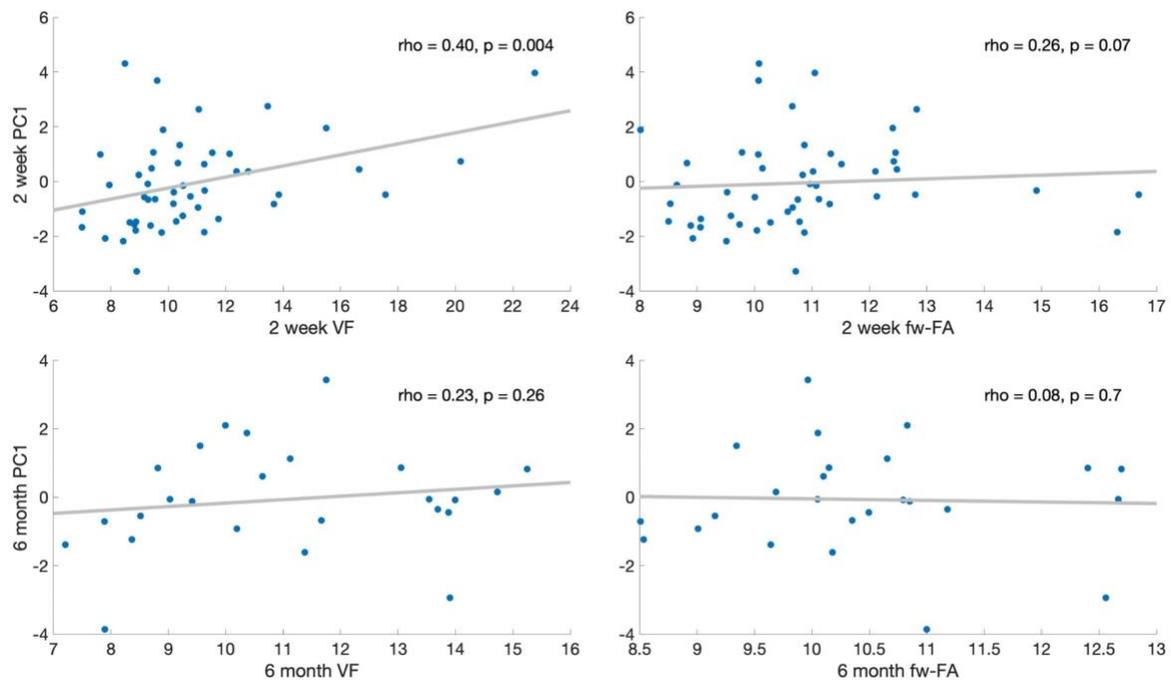